# Modulating the tennis racket grip during motor imagery influences serve accuracy and performance: A pilot study

Aymeric GUILLOT[1], Julien GAUTHIER[1], Jeanne LEJONCOUR[1] & Franck Di RIENZO[1,2]

[1] Inter-University Laboratory of Human Movement Biology-EA 7424, University of Lyon, University Claude Bernard Lyon 1, 69 622 Villeurbanne, France.

[2] Institut Universitaire de France (IUF)

**Email addresses of the authors**: aymeric.guillot@univ-lyon1.fr,

gauthier.julien55@gmail.com, jeanne.lejoncour29@gmail.com, franck.di-rienzo@univ-lyon1.fr

**Correspondence**: Aymeric GUILLOT, Inter-University Laboratory of Human Movement Biology -EA 7424, University Claude Bernard Lyon 1, Villeurbanne, France., E-mail: aymeric.guillot@univ-lyon1.fr

# Modulating the tennis racket grip during motor imagery influences serve accuracy and performance: A pilot study

**Abstract**

There is now ample evidence that Motor Imagery (MI) contributes to improve motor performance. Previous studies provided evidence that its effectiveness remains dependent upon specific guidelines and recommendations. The body posture, as well as the context in which MI is performed, are notably critical and should be carefully considered. The present study in young tennis players (n=18) was designed to compare the effectiveness of performing MI of the serve while adopting a loose grip (congruent MI) or holding tightly and squeezing hard the racket (incongruent MI). Data revealed that both MI conditions contributed to enhance the number of successful serves ($p<0.001$) and the technical quality of the serve ($p<0.001$). Interestingly, comparing mean serve accuracy scores showed that performance gains were significantly higher in the loose MI group than in the tight MI group ($p<0.02$). These findings confirm the critical importance of the congruence between the content of the mental representation and the features of the corresponding actual movement. Overall, the present study further highlights the effectiveness of the loose grip while mentally rehearsing the serve, and might thus contribute to update and adjust specific MI guidelines and recommendations.

# Moduler la prise de raquette pendant l'imagerie motrice influence la précision et la performance du service en tennis : une étude pilote

**Résumé**

La contribution de l'imagerie motrice (IM) à l'amélioration des performances est désormais bien établie, bien qu'assujettie à des préconisations d'usage et consignes de pratique spécifiques. En particulier, la position et le contexte dans lequel l'IM est effectuée exercent une influence sur sa qualité et son efficacité. L'objectif de cette étude chez de jeunes joueurs de tennis (n=18) était de comparer l'efficacité de deux conditions d'imagerie motrice (IM) : une condition compatible où le joueur imagine son service en tenant la raquette de manière relâchée (IM de relâchement), et une condition incompatible où il imagine tenir la raquette fermement pour créer une tension musculaire (IM de contraction). Les données ont montré que les deux conditions d'IM contribuaient à améliorer le taux de réussite ($p<0.001$) et la qualité technique du geste de service ($p<0.001$). En revanche, le score de précision moyen du service était significativement plus élevé après l'IM de relâchement ($p<0.02$), confirmant l'importance de la compatibilité entre contenu des représentations mentales et exigences de la pratique réelle. Ces données soulignent ainsi l'efficacité du relâchement de la main et permettent d'affiner les consignes liées au travail mental chez les jeunes joueurs de tennis.

**Mots-clés** : Entraînement mental ; Relâchement ; Apprentissage moteur.

**Introduction**

Motor imagery (MI) is a multisensory experience during which athletes mentally represent a movement by recalling previously perceived situations or elaborating on forthcoming events. There is now compelling evidence that MI contributes to improve motor performance and facilitate motor learning (e.g., McNeill et al., 2020; Ladda et al., 2021; Simonsmeier et al., 2021). MI remains however subjected to guidelines and rules of practice to maximize effectiveness of training interventions (Guillot, 2020 for an extensive review). Among these recommendations, the degree of compatibility between the imagined movement and the body position with the corresponding movement have been early considered (e.g., Saimpont, Malouin, Toussignant & Jackson, 2012; Saimpont et al., 2021). Accordingly, the existence of tactile and proprioceptive input during MI was found to facilitate corticospinal excitability (Fourkas, Ionta & Aglioti , 2006), in particular when the body position was congruent with that of the actual practice (Vargas et al., 2004), and while a congruent object was concurrently held during MI (Mizuguchi et al., 2012; Mizuguchi, Sakamoto, Murakoa & Kanosue, 2009). Feeling the sensations usually induced by actual execution would thus enhance both accuracy and effectiveness of MI (Holmes & Collins, 2001, Guillot & Collet, 2008; Conson, Mazzarella & Trojano, 2011). As well, holding an object during MI might enhance the activity of the frontoparietal network, known to play a critical role during MI by facilitating simulation of feeling in association with the actual execution (Mizuguchi et al., 2013). A recent study by Ali, Montani and Cesari (2023) reported increased amplitude of motor-evoked potentials of the relevant muscle when imagined actions were performed concurrently with tactile stimulation, hence confirming the positive effect of touch on the motor system. Taken together, these data provided convergent evidence for a congruent form of MI practice reproducing endogenous (physiological arousal) and exogenous (environmental) contexts of the physical performance (an exception to this principle was reported by Kanthack, Guillot, Saboul, Debarnot, and Di Rienzo, 2019, who demonstrated that in rare cases, an incongruent form of MI could still be beneficial, i.e. apneists imagined themselves breathing while actually holding their breath, which paradoxically led to performance improvements).

In this framework, congruent MI refers to a situation where the imagined action matches the usual sensorimotor conditions of its execution, while incongruent MI involves a deliberate mismatch between imagery content and the concurrent bodily or environmental context (Holmes & Collins, 2001; Guillot & Collet, 2008). Although both forms can influence performance, congruent imagery is generally considered more effective, with incongruence sometimes leading to suboptimal or paradoxical effects (Kanthack et al., 2019). In racket sports such as tennis, this distinction can be operationalized through the way players hold their racket during MI. Because grip pressure directly modulates proprioceptive and tactile sensations, it provides a meaningful lever to contrast congruent (e.g., loose grip) and incongruent (e.g. tight grip) conditions, thereby offering an ecological way to examine how sensory congruence impacts serve performance.

Based on these findings, the present pilot experiment aimed to explore and extend these effects in an ecological study involving young elite tennis players, specifically examining whether MI effectiveness is influenced by the way players manipulate tactile and proprioceptive sensations while holding their racket during MI of their serve. To investigate this, we compared the effects of performing MI in two different conditions: one where players adopted a loose grip and another where they held the racket tightly, applying strong pressure and squeezing hard the racket. We postulated greater effects for the loose grip condition, due to a decreased tension in the forearm during the serve. While we acknowledge that tightly gripping the racket is not a common practice during the ball toss in actual play, this condition was designed to assess the influence of different pre-conditioning states induced by MI rather than to replicate a real-life serving scenario. We hypothesized that the loose grip condition would yield greater benefits due to reduced forearm tension, potentially facilitating a more fluid and efficient serve execution.

**Methods**

*Participants*

Eighteen young tennis players (3 women and 15 men, mean age 15.40 ± 2.46 years), with more than 9 years of practice and training more than 7h per week, volunteered to participate in this study (see Table 1 for detailed information in each group). All players were high-level regional players athletes selected from their regional tennis league's elite training group. Before the experiment, they completed the third version of the Movement Imagery Questionnaire – French translation (MIQ-3f; Robin, Coudevylle, Guillot & Toussaint, 2020). Written informed consent was obtained from all participants and parents before data collection, in accordance with the Declaration of Helsinki. The study was scheduled during their regular training sessions.

*Table 1. Individual characteristics of the participants in each MI group (Mean ± SD).*

|  | Age | Height (cm) | Years of practice |
|---|---|---|---|
| Loose motor imagery group | 15.20 ± 1.81 | 1168.54 ± 9.02 | 9.45 ± 2.51 |
| Tight motor imagery group | 17.00 ± 2.61 | 173.4 ± 6.10 | 10.2 ± 2.57 |

*Experimental design*

The effects of a 5-week MI training intervention were evaluated in a test-retest experimental design. Tennis serve performance was assessed before and after the intervention by having participants complete a standardized serve accuracy test. Each player performed 16 consecutive serves (8 from the right side and 8 from the left side of the court) to ensure balanced evaluation across diagonals. Players were instructed to hit serves as fast and accurately as possible within a predetermined target zone, simulating the intention of hitting an ace in a competitive match (Figure 1).

*Serve Accuracy Scoring System*: The target was based on the 'T' of the court, located at the intersection of the service-box and center lines. This zone was chosen because it corresponds to the lowest point of the net, making it a strategic and mechanically advantageous target compared to a wide cross-court serve. Targeting this area therefore reduced the risk of net faults and facilitated controlled accuracy, allowing participants to focus on precision while maintaining realistic serving conditions. The target zone was divided into three scoring zones to quantify accuracy, following a procedure adapted from Guillot, Deslien, Rouyer, and Rogowski (2013).

- ✓ Small target area (0.5 × 0.5 m) accounted for five points.
- ✓ Medium target area (1 × 1 m) accounted for three points.
- ✓ General service box (outside the two smaller zones) accounted for one point.
- ✓ Any other location (outside the service box) yielded zero point (fault).

*Serve success rate*: the percentage of successful serves landing in the service box, regardless of the scoring zone.

*Serve accuracy score*: the sum of all points obtained over the 16 serves, providing a more fine-grained measure of precision.

All serves were recorded using two synchronized video cameras, capturing different angles for precise post-analysis. The first camera was positioned 2 meters behind the server, at a height of 1.5 meters, aligned with the court centerline. This perspective (frontal baseline view) provided a direct view of the ball toss, body posture, and follow-through mechanics. The second camera was placed 3 meters to the side of the server, aligned with the server, at a height of 1.2 meters. This angle (lateral view) allowed for detailed assessment of racket trajectory, timing of contact, and weight transfer. A tennis expert with over 10 years of coaching experience at the national level independently analyzed the video recordings. The expert used a validated 30-item technical evaluation questionnaire (Guillot, Genevois, Desliens, Saieb & Rogowski, 2012) to rate the quality of each serve. Examples of key elements analyzed included: Ball toss consistency (height, positioning, and timing relative to the striking motion), body coordination and balance (weight transfer, trunk rotation, and synchronization of movements), racket acceleration and contact timing (smoothness of the motion and energy transfer), or follow-through mechanics (fluidity and completion of the serve motion). Each technical aspect was scored on a 10-point scale, ranging from 0 (poorest execution) to 10 (best execution). To ensure objectivity and prevent bias, the expert was blinded to the study's hypotheses and the participants' group allocation (loose vs. tight grip MI groups). To ensure standardized viewing conditions, all videos were analyzed in a quiet environment on a high-resolution screen, allowing frame-by-frame review when necessary. The total evaluation process lasted approximately 30 to 45 minutes per participant, depending on the need for replay and detailed technical analysis.

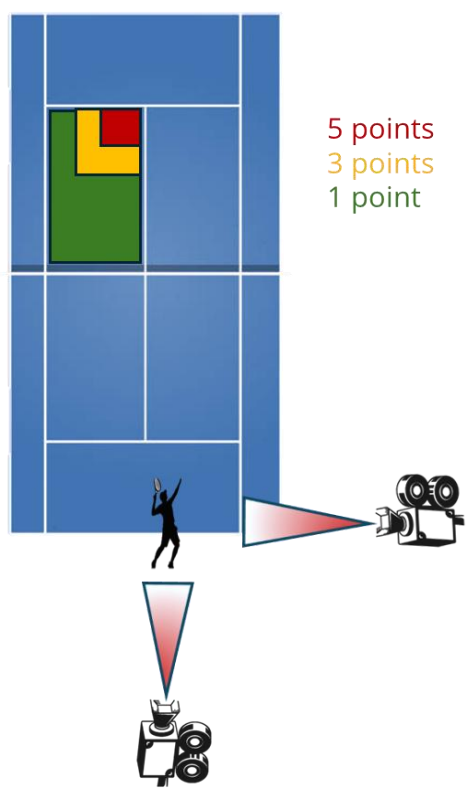

Figure 1. Testing tennis court instrumentation. In this example, the serve is taken from the right diagonal position.

*Warm-up procedure*: Before starting the serve accuracy test, all participants followed a standardized warm-up routine to ensure optimal preparation and minimize injury risk. The warm-up consisted of 5 minutes of general dynamic warm-up (e.g., jogging, arm swings, and shoulder mobility exercises), and 5 minutes of specific tennis warm-up, including shadow swings and submaximal serves to progressively increase serve intensity. Participants were instructed to gradually increase their serve speed during the warm-up to ensure that their first recorded serve was performed at their maximum intensity.

During the five weeks of the protocol, specific training sessions including MI exercises were administered twice per week. Participants were randomly assigned either to the *loose grip* (n=9) or the *tight grip* (n=9) MI group. Practically, players mentally rehearsed one serve before subsequent physical practice trial. Each session included 20 imagined and 20 actual tennis serves, presented in a counterbalanced order (both diagonals per session), with the same instructions than during the testing protocol. An imagery script including first-person visual perspective as well as kinesthetic and tactile imagery was read to the participants at the beginning of each MI session to ensure that they received comparable imagery instructions. The only difference was related to the grip of the dominant hand holding the racket. The loose grip MI group was asked to maintain a relaxed grip during the ball toss and the serve, while the tight grip MI group was requested to firmly hold and hardly squeeze the racket throughout the motor sequence. Every two MI sessions (i.e. at sessions 2, 4, 6, 8 and 10), participants reported their perceived MI vividness on a 6-point Likert-type scale (1: "*Absence of sensations/images, only thinking about the movement*"; 6: "*Identical sensations/images as during physical practice of the task*").

      To ensure that participants fully understood the distinction between the loose and tight grip conditions and correctly applied the instructions during MI practice, several control measures were implemented. First, before the start of the training intervention, each participant received a detailed explanation of the two grip conditions. To reinforce understanding, experimenters provided practical demonstrations of the loose and tight grips using a racket, highlighting the differences in hand tension and proprioceptive feedback. Second, participants were questioned individually before the first session using open-ended questions to confirm that they clearly differentiated between the two grips and could accurately describe the sensations they were expected to imagine during MI. Throughout the training sessions, experimenters systematically debriefed about participants' execution of MI and physical trials, ensuring adherence to the grip instructions. If necessary, verbal reminders were provided to reinforce correct application. Finally, players were regularly asked to self-report their ability to mentally rehearse the serve while adopting the prescribed grip condition.

4By implementing these measures, we ensured that participants clearly understood and correctly applied the grip instructions during both MI and physical practice.

Furthermore, to clarify the nature of the grip manipulation, participants were explicitly instructed to maintain their habitual grip type (e.g., continental grip or hammer grip) throughout the study. The experimental conditions were designed to modify only the grip pressure rather than the grip configuration itself. In the loose grip condition, participants were asked to relax their grip while maintaining their natural hand positioning on the racket handle. In the tight grip condition, they were instructed to firmly squeeze the racket while keeping the same grip placement. This approach ensured that the experimental manipulation focused solely on grip pressure modulation, allowing us to isolate the effects of a more relaxed versus a more forceful grip during MI without altering the biomechanical structure of the serve.

To summarize and provide a clear overview of the study design, the independent variables were the MI conditions, with both the loose grip and the tight grip MI groups, while the dependent variables used to assess the effects of MI training included serve accuracy metrics (percentage of successful serves, serve accuracy score, technical execution quality, evaluated by a blinded expert, and MI vividness scores).

*Data analyses*

Data are presented as Mean ± Standard Deviations. The normality of the data was checked and confirmed that parametrical statistical tests could be used despite the small sample size. An analysis of variance was used to compare MIQ-3f scores as well as MI vividness scores among groups, and analyses of variance for repeated measures were performed to compare the effects of training. F- and p-values, as well as partial effect sizes ($\eta_p^2$), are reported. The level of statistical significance was set at $p \leq 0.05$.

**Results**

*Tennis serve accuracy*

Data analysis revealed a main TEST effect (F(1,16) = 41.09; p < 0.001, $\eta_p^2$ = 0.74) as well as a statistically significant GROUP × TEST interaction for the serve accuracy (F(1,16) = 6.38; p = 0.02, $\eta_p^2$ = 0.28). In the loose MI group, the mean accuracy score increased from 13.22 ± 5.19 during the pre-test to 20.89 ± 4.88 during the post-test. In the tight MI group, the mean accuracy score increased from 13.44 ± 5.50 during the pre-test to 16.78 ± 6.02 during the post-test (Figure 2).

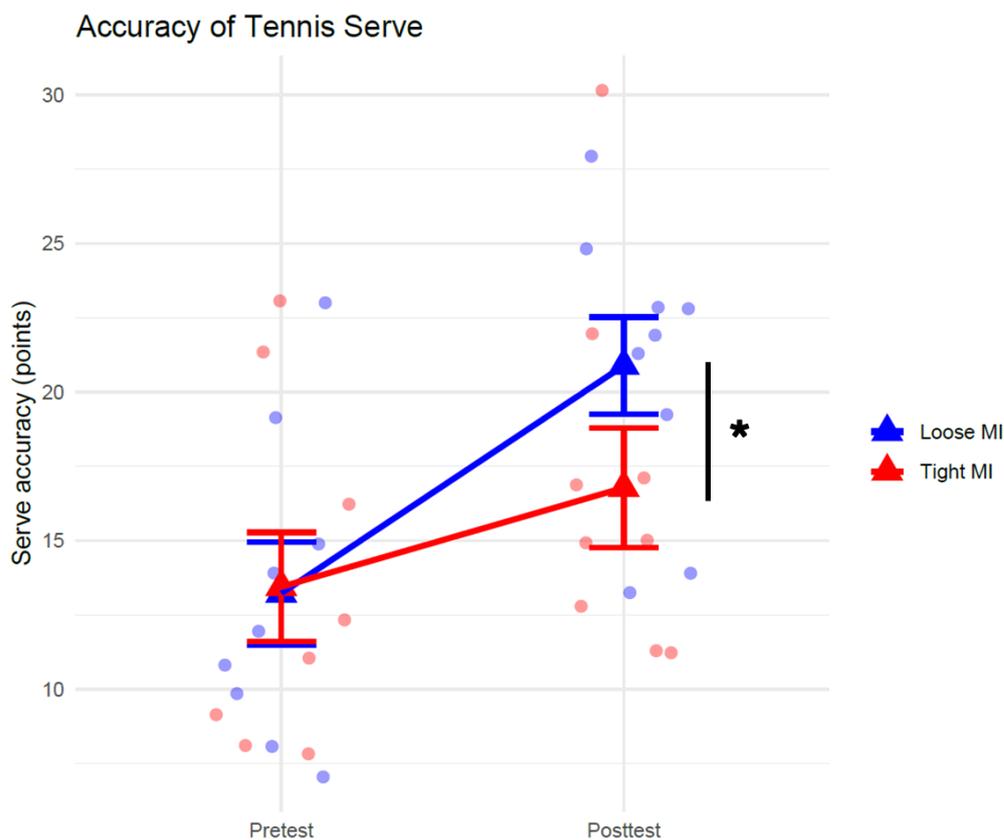

Figure 2. Serve accuracy

The ANOVA showed a main TEST effect for the percentage of successful serves (F(1,16) = 21.3; p < 0.001, $\eta_p^2$ = 0.57; Figure 2). During the pre-test, the mean percentage of successful serves was 46.53 % ± 10.42 in the loose and 40.28 % ± 13.30 in the tight MI group, respectively. Respective mean percentages were 61.11 % ± 9.26 in the loose and 50.69 % ± 13.05 in the tight MI group during the post-test (Figure 3A). There was no GROUP × Test interaction (F(1,16) = 0.59, p = 0.45, $\eta_p^2$ = 0.04).

For the serve technical quality, ANOVA did not show a GROUP × TEST interaction (F(1,16) = 0.46, p = 0.51, $\eta_p^2$ = 0.05), but technical serve quality was affected by the main TEST effect (F(1,16) = 31.2; p < 0.001, $\eta_p^2$ = 0.66; Figure 2). During the pre-test, the mean technical quality rated by the expert coach was 242.67 ± 18.90 in the loose and 238.44 ± 25.21 in the tight MI group, respectively. Respective mean scores were 261.44 ± 8.44 in the loose and 262.67 ± 16.09 in the tight MI group during the post-test (Figure 3B).

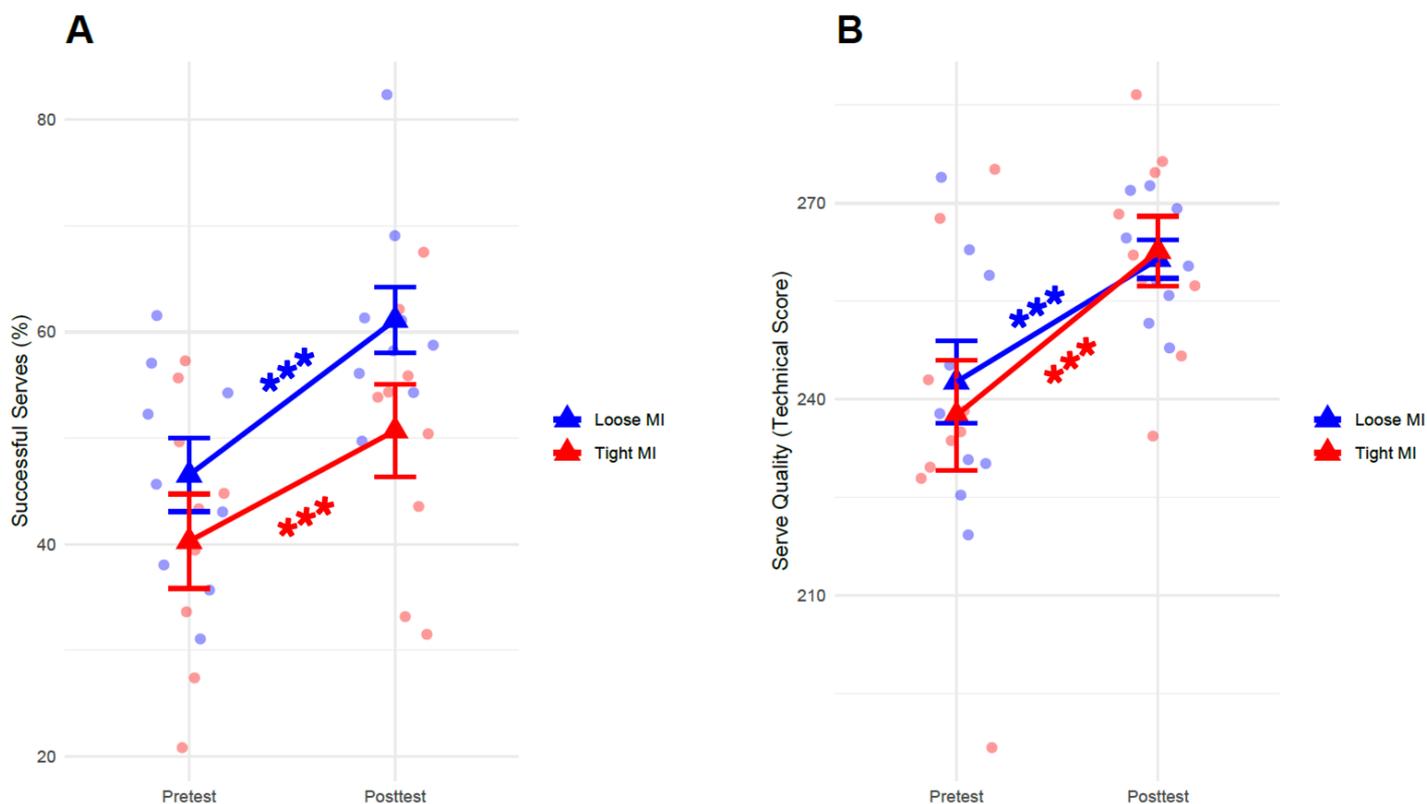

Figure 3. Successful serves (A) and serve quality (B)

*Motor imagery vividness*

There was no GROUP difference when comparing mean MIQ-3f scores ($F(1,16) = 0.07$, $p = 0.79$, $\eta_p^2 = 0.004$), which were $5.06 \pm 0.76$ in the loose MI group, and $4.95 \pm 0.91$ in the tight MI group. Also, there was no main effect of MODALITY nor GROUP × MODALITY interaction (all $p > 0.05$).

MI vividness scores were unaffected by the GROUP × SESSION interaction ($F(4,64) = 0.48$, $p = 0.75$, $\eta_p^2 = 0.07$). There was however a main GROUP effect ($F(1,16) = 5.54$, $p = 0.03$, $\eta_p^2 = 0.26$), respective mean scores being $4.33 \pm 1.07$ in the loose and $3.38 \pm 0.58$ in the tight MI group (Table2). A main SESSION effect was also observed ($F(4.64) = 6.83$, $p < 0.001$, $\eta_p^2 = 0.30$). Post-hoc tests with Bonferroni corrections revealed that the MI vividness scores reported during the last session were higher than those reported during the first three sessions (all $p < 0.01$).

*Table 2. Vividness motor imagery score for all time points across training*

|  |  | Session 2 | Session 4 | Session 6 | Session 8 | Session 10 | Mean score |
|---|---|---|---|---|---|---|---|
| Loose MI group | Mean | 3.78 | 3.89 | 4.56 | 4.44 | 5.00 | 4.33 |
|  | SD | *1.30* | *1.27* | *0.73* | *1.59* | *1.12* | *1.07* |
| Tight MI group | Mean | 3.22 | 3.00 | 3.22 | 3.67 | 3.78 | 3.38 |
|  | SD | *0.67* | *1.00* | *0.83* | *0.71* | *0.67* | *0.58* |

**Discussion**

The present study was designed to investigate whether adopting a loose or a tight tennis racket grip during MI was likely to affect subsequent tennis serve performance. Data confirmed previous findings supporting the greater effectiveness of a congruent form of MI with the corresponding physical execution of the task (Vargas et al., 2004), as well as the related compatibility of the body posture (Saimpont et al., 2012). Specifically, this form of MI would more closely reflect the features of the motor task, where the athlete usually looks for reduced tension and easy fluid motion. Subjective self-reported MI vividness evaluations were on average higher in the loose (congruent) form of MI and increased across sessions in both groups, although no Group × Session interaction was observed. While the beneficial effects of MI on motor performance are well-established, including for tennis serve performance, previous work primarily focused on the content of the MI training program such as the efficacy of the external attentional focus (Guillot et al., 2013), or the effect of self-talk during MI (Robin et al., 2021). The present study went further by investigating, in an ecological context, the effect of different tactile and proprioceptive inputs induced by concurrently holding the object used during the physical practice of the same movement. As suggested by Mizuguchi et al. (2009, 2012, 2013) and Ali et al. (2023), holding the racket facilitated the mental representation of tennis actions. It contributed to generate closer sensations during mental rehearsal to those felt during the actual movement, which resulted in improved subsequent tennis serve accuracy.

      The grip is a success factor in tennis (Elliott, Marshall & Noffal, 1995). From a biomechanical approach, the loose grip helps to manipulate the tennis racquet more efficiently, allowing for a more fluid and relaxed motion which can generate higher racket head speed (Elliott, Takahashi & Noffal, 1997). With a loose grip, athletes could also more easily pronate their wrist during the serve, thus creating spin on the ball, and further reduce the tension and the stress on the arm. More specifically, we postulate that a looser grip may enhance proprioceptive and kinesthetic feedback, improving the sensorimotor representation of the serve during MI. Research suggests that overly rigid grip pressure may reduce tactile sensitivity, potentially diminishing the effectiveness of mental simulation by limiting the perception of fine movement details. By contrast, using a loose grip is likely to limit serve

power (Hatze, 1976), and may reduce racket stability and control (Knudson & Elliott, 2004), due to greater instability at the point of contact. To compensate for this lack of stability, players might instinctively increase muscular tension in other parts of their body, such as the arm, which could lead to a loss of fluidity and a potential decline in overall performance. As well, a tight grip may introduce unnecessary muscle co-contractions involving proximal muscle groups such as the shoulder or trunk. This increased muscular rigidity could interfere with efficient energy transfer and fluid coordination, essential for an effective serve. As a result, a loose grip might facilitate a more natural and biomechanically efficient execution, both in imagery and in actual movement.

Despite these biomechanical considerations, our results revealed the superiority of the loose grip over the tight grip during MI in terms of serve accuracy improvement. While both groups showed enhanced performance from pre- to post-test, the improvement was significantly greater in the loose grip group. This suggests that the excessive voluntary contraction in the tight grip condition may have induced an unnatural tension in the forearm, potentially disrupting the fluidity of the imagined movement and reducing the efficiency of the mental rehearsal process. Although the tight grip condition still led to a measurable improvement in accuracy, this increase was more limited compared to the loose grip condition, reinforcing the idea that a more relaxed MI state better aligns with the actual execution of a skilled movement. This finding further supports the notion that MI effectiveness is modulated by the congruence between the imagined and the executed movement, particularly regarding proprioceptive and tactile sensations.

Interestingly, data revealed that only the loose grip MI reached a serve accuracy score reflecting a substantial reduction in faults and missed serves (i.e., a score superior to 20 points). This result indirectly confirms that MI with a relaxed grip may have enhanced both execution consistency and overall control, potentially by reducing unnecessary muscle tension and allowing for a more natural and fluid serve motion. The tight grip MI condition, in contrast, may have introduced excessive rigidity, limiting fine motor adjustments necessary for precise targeting, hence resulting in a higher number of failed serves.

Data arising from the self-reported imagery vividness evaluations further suggest that maintaining a loose grip during MI was, on average, associated with higher vividness ratings across sessions, potentially due to a greater sense of fluidity and reduced muscular tension. Conversely, the tight grip condition may, once again, have introduced unnecessary physical tension, making it more challenging for participants to engage in a detailed and immersive imagery experience. However, since vividness increased in both groups and no Group × Session interaction was observed, differences in vividness alone cannot fully explain the superior performance of the loose group. Instead, these findings strengthen the interpretation that proprioceptive congruence plays a key role in optimizing MI effectiveness. As baseline MI vividness was comparable between groups, meaning that initial imagery ability was not a confounding factor, we interpret the performance differences as more directly reflecting the influence of grip pressure on the quality and functional relevance of mental rehearsal, rather than changes in vividness across training.

*Limitations*

Practically, the present study shed light on the optimal conditions of MI practice when the aim is to enhance tennis serve performance, demonstrating the superiority of the loose grip over the tight grip. This nonetheless remains a pilot study, and several limitations must be acknowledged before drawing firm conclusions. First, the study design does not allow us to firmly determine whether the performance improvement observed in the loose grip group was due to a specific enhancement effect or whether the tight grip condition interfered with motor execution. While both groups showed an increase in performance, the absence of a pure control group performing either neutral MI or physical practice alone makes it difficult to isolate the selective influence of each MI condition. This omission makes it difficult to determine whether the observed improvements are specifically due to MI or simply the result of repeated serve practice over the five-week period. Future studies should thus incorporate additional experimental conditions to clarify whether congruent MI optimally enhances performance or whether incongruent MI disrupts it, or both, as well as pure control groups performing only MI or physical practice.

Another main limitation is the lack of controlled serve speed measurements. Although players were instructed to serve as fast as possible, as if they were serving for an ace, controlling for and analyzing serve speed might have provided further insights into how MI conditions influenced both accuracy and power. While players were explicitly instructed to serve as fast as possible, we cannot fully exclude the possibility that some participants subconsciously adjusted their serve speed to improve accuracy, leading to a potential speed-accuracy trade-off. Originally, serve speed was intended to be recorded in this study, but technical issues and inconsistencies in data collection prevented us from incorporating this variable into our analysis. As a result, we were unable to determine whether changes in accuracy were accompanied by compensatory adjustments in speed. Future research should thus integrate precise and reliable speed-tracking methods serve speed data (e.g., radar measurement or high-speed motion capture) to evaluate the potential trade-off between serve velocity and accuracy and determine whether MI conditions influence this relationship.

Finally, a major common limitation is the lack of strict control over participants' additional training outside of the experimental sessions. While we took measures to minimize this potential confound by informing players' coaches and asking them to avoid specific training related to serves during the five-week period, we cannot fully guarantee that participants did not engage in additional unsupervised practice. To further mitigate this issue, we relied on regular self-reports from players, confirming that they did not train independently beyond the study protocol. However, as these reports were based on verbal statements rather than objective tracking, we acknowledge that some uncontrolled variability may have been present. Nonetheless, because players served as their own controls in the experimental design, any significant extra training should have been detectable in the dataset as an unusual improvement pattern. Given the consistency of the results across participants, we believe that any potential out-of-protocol practice had a limited impact on our findings.

*Perspectives*

Despite the methodological limitations discussed above, the present study presents several key strengths and paves the way for insightful use of MI in tennis.

While future studies should examine the role of proprioceptive and tactile input during MI across a larger and more diverse sample of players including different skill levels and age groups, the present work investigated the benefits of MI in a population of young athletes that usually remains underrepresented in MI research. MI has been primarily and extensively studied in adult athletes, and there is still a lack of empirical data on its effectiveness in young developing players, despite growing evidence suggesting that MI is a skill that matures throughout childhood and adolescence (for review, see Guilbert et al. 2013; Guilbert & Fernandez, 2023). By investigating the effects of MI in this population, our study contributes to a better understanding of how imagery-based interventions can be integrated into youth training programs. Given that young athletes are in a critical period for motor learning and skill refinement, the ability to optimize MI techniques from an early stage could enhance long-term skill acquisition and competitive performance. Future research should further explore developmental aspects of MI, examining how factors such as age, cognitive maturation, and experience level influence imagery ability and its impact on motor performance in young athletes.

Another strength of the present study is our effort to maintain ecological validity by testing players in a real serve situation after different MI conditions. While our protocol focused on isolating specific variables, it also opens promising avenues for future research aiming to approximate match conditions even more closely. For instance, investigating how the presence of an opponent, the two-serve rule, or the continuation of points might affect MI efficiency and serving behavior would enrich our understanding of MI in competitive settings. Moreover, integrating psychological factors such as score-related pressure (e.g., serving under break point conditions) could provide valuable insights into how competitive context modulates grip dynamics and tension regulation. Exploring these directions would significantly enhance the transferability of MI training to real match play.

A final key strength of this study is its focus on the mental representation of racket grip during MI, an aspect of MI that has received little attention in the literature. While previous studies have explored the impact of MI on global movement execution, few have specifically examined how proprioceptive and tactile sensations related to grip pressure influence mental rehearsal and subsequent motor performance. By manipulating grip pressure as an experimental factor, our study therefore provides new insights into how fine motor control and preparatory muscle tension interact with MI processes. These findings confirm the importance of sensory congruence in MI practice, suggesting that pre-conditioning grip pressure before MI may influence motor performance. Future research should now explore how different grip configurations, pressures, and even racket weights impact the effectiveness of MI, with potential applications for both technical training and injury prevention in racket sports. Accordingly, while we compared loose and the grips as two 'extreme' opposite conditions, the reality should certainly be seen as a grip continuum ranging from the loose *to* the tight grip. Optimal grip pressure may vary between players, and finding the right balance between grip tension and control remains crucial. Future studies should thus explore the impact of different intermediate grip pressures and personalized MI strategies that align with individual players' biomechanical and perceptual preferences

23Mizuguchi, N., Nakata, H., Hayashi, T., Sakamoto, M., Muraoka, T., Uchida, Y. & Kanosue, K. (2013). Brain activity during motor imagery of an action with an object: a functional magnetic resonance imaging study. *Neuroscience Research*, 76(3), 150-155.

Mizuguchi, N., Sakamoto, M., Muraoka, T., & Kanosue, K. (2009). Influence of touching an object on corticospinal excitability during motor imagery. *Experimental Brain Research*, 196, 529-535.

Mizuguchi, N., Sakamoto, M., Muraoka, T., Moriyama, N., Nakagawa, K., Nakata, H., & Kanosue, K. (2012). Influence of somatosensory input on corticospinal excitability during motor imagery. *Neuroscience Letters*, 127-130.

Robin, N., Coudevylle, G., Guillot, A., & Toussaint, L. (2020). French Translation and validation of the Movement Imagery Questionnaire–third version (MIQ-3f). *Movement and Sport Sciences – Science et Motricité*, 108, 23-31.

Saimpont, A., Malouin, F., Tousignant, B., & Jackson, P.L. (2012). The influence of body configuration on motor imagery of walking in younger and older adults. *Neuroscience*, 222, 49-57.

Saimpont, A., Malouin, F., Durand, A., Di Rienzo, F., Saruco, E., Collet, C., Guillot, A., & Jackson, P. (2021). Effects of current body configuration and movement execution on motor imagery of locomotor tasks in persons with a lower-limb amputation. *Scientific Reports*, 11, 13788.

Simonsmeier B.A., Andronie M., Buecker S. & Frank C. (2021). The effects of imagery interventions in sports: A meta-analysis. *International Review of Sport and Exercise Psychology*, 14(1), 186-207.

Vargas, C.D., Olivier, E., Craighero, L., Fadiga, L., Duhamel, J.R., & Sirigu, A. (2004). The influence of hand posture on corticospinal excitability during motor imagery: a transcranial magnetic stimulation study. *Cerebral Cortex*, 14(11), 1200-1206.